\documentclass[twocolumn]{aastex631}
\usepackage{natbib}
\usepackage{graphicx}
\usepackage{mathrsfs}
\usepackage{amsmath}
\usepackage{tikz}
\usepackage{hyperref}
\usepackage{subfigure}

\tikzstyle{blue} = [rectangle, rounded corners, minimum width=3cm, minimum height=1cm,text centered, draw=black, fill=blue!30]
\tikzstyle{red} = [rectangle, rounded corners, minimum width=3cm, minimum height=1cm,text centered, draw=black, fill=red!30]
\tikzstyle{green} = [rectangle, rounded corners, minimum width=1.5cm, minimum height=1cm,text centered, draw=black, fill=green!30]
\tikzstyle{orange} = [rectangle, rounded corners, minimum width=1.5cm, minimum height=1cm,text centered, draw=black, fill=orange!30]
\tikzstyle{arrow} = [thick,->,>=stealth]

\begin{document}

\title{Quantifying the Effect of Short-timescale Stellar Activity Upon Transit Detection in M Dwarfs}

\author{Dana Clarice Yaptangco}

\author{Sarah Ballard}

\author{Jason Dittmann}

\affiliation{Department of Astronomy, University of Florida, Gainesville, FL 32611, USA}

\begin{abstract}
M dwarf stars comprise 70-80\% of the galaxy’s stars and host most of its rocky planets. They also importantly differ from Sunlike stars in that they are “active” for billions of years or more: rotating quickly, flaring often, and emitting large amounts of UV and X-ray light. The effects of stellar activity upon both photometry and spectroscopy make their exoplanets more difficult to detect, and M dwarfs exhibit this behavior for thousands of times longer than a typical Sunlike star. While activity signals such as flaring and stellar rotation can be more readily modeled or removed from photometry, the contribution of unresolved stellar activity to transit sensitivity is harder to quantify. In this paper, we investigate the difference in the detectability of planetary transits around a sample of M dwarfs observed by NASA's \textit{TESS} Mission, characterized by a common stellar radius, effective temperature, and \textit{TESS} magnitude. Our sample is classified as either ``active" or ``inactive" based upon the presence of H$\alpha$ in emission. After removing the more readily identifiable signatures of activity: stellar rotation and large flares, we perform an injection-and-recovery analysis of transits for each star. We extract detection sensitivity as a function of planetary radius and orbital period for each star in the sample. Then, we produce averaged sensitivity maps for the ``active" stars and the ``inactive" stars, for the sake of comparison. We quantify the extent to which signal-to-noise is degraded for transit detection, when comparing an active star to an inactive star of the same temperature and apparent brightness. We aim for these sensitivity maps to be useful to the exoplanet community in future M dwarf occurrence rate studies. 

\end{abstract}


\section{Introduction}
While the Solar System has historically been the default blueprint for a planetary system in our galaxy, it is atypical in the Milky Way. M dwarfs, in reality, comprise 70-80\% of the galaxy’s stars and host most of its rocky planets \citep{Henry04, Dressing15, Mulders15, Hardegree19, He20, Ment23}. M dwarf stars also differ from Sun-like stars in that they are “active” for billions of years or more \citep{West08, Newton2016}. Decades of research have elucidated the physical processes linked to stellar “activity," a term used to describe a range of phenomena associated with the strength of the star’s magnetic field: X-ray emission, H$\alpha$ in emission, rotation periods generally inferred from photometric monitoring, and starspot and faculae coverage \citep{Giampapa86, Soderblom91, Reid95, Hawley96, Delfosse98, Hawley99, Barnes2004, Oneal04,  Irwin11, Newton17}. Below the convective limit, there is evidence for two divergent populations in the M spectral class (see e.g. \citealt{Irwin11, Newton17, Popinchalk21, Kiman21}): ``active" faster rotators exhibiting H$\alpha$ in emission, and slower/``inactive" rotators exhibiting H$\alpha$ in absorption. This gap is likely attributable to a period of short but rapid spin-down of mid-to-late M dwarfs. However, measuring a slow/``inactive" rotator still requires stellar starspots on its surface to rotate in and out of view, so even an inactive rotator is not completely inactive. 

Stellar activity contributes noise to high-precision photometric observations in ways that are varied and complex. Some, like stellar rotation, spot evolution, and large flares, can be modeled and removed (see e.g. \citealt{Aigrain15, Davenport16, Hawley14}), although stellar rotation periods at the same period as potential orbiting planets can complicate detection. In other ways, the identification of the effects of stellar activity upon lightcurves is ongoing: photometric ``flicker" on $\sim$8-hour timescales was only detected in 2013, for example \citep{Bastien13, Bastien16}. The effects of pulsations, granulation, long-term evolution of active regions, and magnetic cycles are also present, but can be difficult to resolve and model. Several tools including SOAP \citep{Boisse12, Dumusque14} and StarSim \citep{Herrero16} exist for ameliorating these effects, often by modeling photometry and radial velocity in tandem. Because timescales associated with stellar activity range from minutes to years \citep{Medina22, Basri13, Suarez16, Mignon23}, resolving the signatures of activity in a given lightcurve depends upon its precision, cadence, duty cycle, and duration. 


Because of their abundance in nature, it is of interest to investigate the sensitivity of transit surveys to exoplanets around M dwarf stars. Across the M spectral class, the ``active" duration of a star's life varies from 1 Gyr in the case of M0 dwarfs to 8 Gyr or more for spectral type M8 \citep{West08, West15}; by spectral type M6, at least half of field stars are active (though this is also dependent upon galactic radius and height, per \citealt{Pineda13}). The relevance to transit searches is strong: active stars present photometric variability over a range of timescales, but crucially, some are extremely close to the transit durations of transiting planets. A Gaussian process regression of photometry of Proxima Centauri by \cite{Kipping17} showed a characteristic correlation length of 130 minutes. \cite{Medina22} identified that the characteristic variability of H$\alpha$ emission of 20-25 minutes corresponded to variability on that timescale in photometry as well. A study of the hot Jupiter HD 189733b by \citet{Cauley2017} found that the depth of its transit in the H$\alpha$ line varied between transits over the host star's activity cycle. Additionally, this variability did not correlate significantly with traditional activity indicators like the Ca H\&K lines, suggesting that decorrelating planet signals from stellar activity signals will require more complicated approaches. If  H$\alpha$ variability is tied to photometric variability in active stars, we can expect correlated noise timescales of 15m-1hr for late-type dwarfs \citep{Kruse10, Bell12, GarciaSoto2023}. Indeed, \cite{Davenport16} estimate that Proxima Centauri exhibits a 0.5\% brightness increase once every $\sim$20 minutes (on average). This in in comparison with an average transit duration of 2.5 hours among the M dwarf exoplanets identified by the \textit{Kepler} Mission \citep{Swift15}. 

We describe here an empirical analysis to characterize the effect of shorter-timescale stellar activity in late-type M dwarfs upon photometric transit detection. While each star's activity signature will be unique, we aim to quantify the extent to which sensitivity to transits degrades \textit{on average}. We design an experiment by which we identify a sample of M dwarfs of common temperature, radius, and apparent magnitude, all observed by NASA's Transiting Exoplanet Survey Satellite \citep{Ricker14}. We then split the sample into ``active" and ``inactive" M dwarfs, using the the presence of H$\alpha$ in emission from \cite{Newton17} as our criterion. We remove the more readily-identifiable signatures of activity: stellar rotation and flares. We then conduct an injection-and-recovery analysis of transit signals for each lightcurve, varying planetary orbital period and radius. Finally, we investigate the extent to which the ability to recover transit signals is affected, comparing the resulting sensitivity maps from the active and inactive samples. 

This manuscript is organized as follows. In Section \ref{sec:Methods}, we describe our sample selection. We detail our methodology for injecting transit signals and then detrending the lightcurves, which we apply to the active and inactive lightcurves in a uniform way. We then attempt to blindly recover the injected signals using publicly available transit search packages. In Section \ref{sec:analysis} we examine the difference in transit sensitivity between the two samples. We quantify, based on averaged sensitivity maps between a fiducial ``active" and ``inactive" star, the effect to which transit signal-to-noise is degraded in the photometry of active stars, even when obvious signs of activity are removed. In Section \ref{sec:conclusion}, we summarize our findings and conclude. 

\section{Methods}
\label{sec:Methods}
To determine the detectability of transiting planets orbiting active versus inactive M dwarfs, we first select a sample of stars with extant transit photometry from the Transiting Exoplanet Survey Satellite (hereafter \textit{TESS}, \citealt{Ricker14}). We describe this procedure in Section \ref{sec:stellar_sample}. With the sample selected, we then detrend the lightcurves to remove signatures of stellar rotation and flaring. We detail the detrending process, which we apply in the same way to both active and inactive stars, in Section \ref{sec:detrending}. In order to explore the parameter space of interest, we conduct injection-and-recovery tests for planetary transits across a range of orbital period and radius. We lay out our framework for conducting this injection-and-recovery analysis in Section \ref{sec:injection}. 

\subsection{Stellar Sample}
\label{sec:stellar_sample}
We draw the stellar sample from among the 270 nearby M dwarfs observed in optical spectroscopy by \cite{Newton17}. That work identifies the presence of H$\alpha$ in emission or absorption for the sample, in addition to noting the near-universality of  H$\alpha$ emission among rapidly rotating stars. There are many signifiers of stellar activity, and here we adopt H$\alpha$ in emission as our diagnostic criterion. From these stars, we identify those with an associated \textit{TESS} Input Catalog \citep{Muirhead17, Stassun19} ID within our dataset, and then remove those with a binary companion, also per \citealt{Newton17}. We also ensure that none of the stars in our sample host a \textit{bona fide} \textit{TESS} Object of Interest \citep{Guerrero}.

\begin{table}[ht]
  \hspace{-.4cm}
  \begin{tabular}{ccc}
    \toprule
    \textbf{TIC ID} & \textbf{TESS Sector} & \textbf{Activity Flag} \\
    \hline
    347695698 & 10 & 1 \\
    238865036 & 59 & 1 \\
    29168887 & 21 & 1 \\
    233068870 & 14 & 1 \\
    157853745 & 14 & 1 \\
    471012250 & 28 & 1 \\
    35760711 & 4 & 1 \\
    297961454 & 21 & 1 \\
    137025855 & 15 & 1 \\
    359185852 & 23 & 1 \\
    274626553 & 24 & 0 \\
    340807280 & 55 & 0 \\
    115131636 & 19 & 0 \\
    446098056 & 19 & 0 \\
    224269420 & 2 & 0 \\
    239481097 & 54 & 0 \\
    80070067 & 43 & 0 \\
    345874162 & 25 & 0 \\
    286582903 & 26 & 0 \\
    188640773 & 24 & 0 \\
  \end{tabular}
    \caption{\textit{TESS} Input Catalog IDs and Sectors of the 20 stars in our sample, with corresponding activity flags drawn from \cite{Newton17}. Here 1 is ``active", signifying H$\alpha$ in emission, and 0 is inactive.}
    \label{tab:sample}
\end{table}

\begin{figure}[h]
  \centering
  \includegraphics[width=1\linewidth]{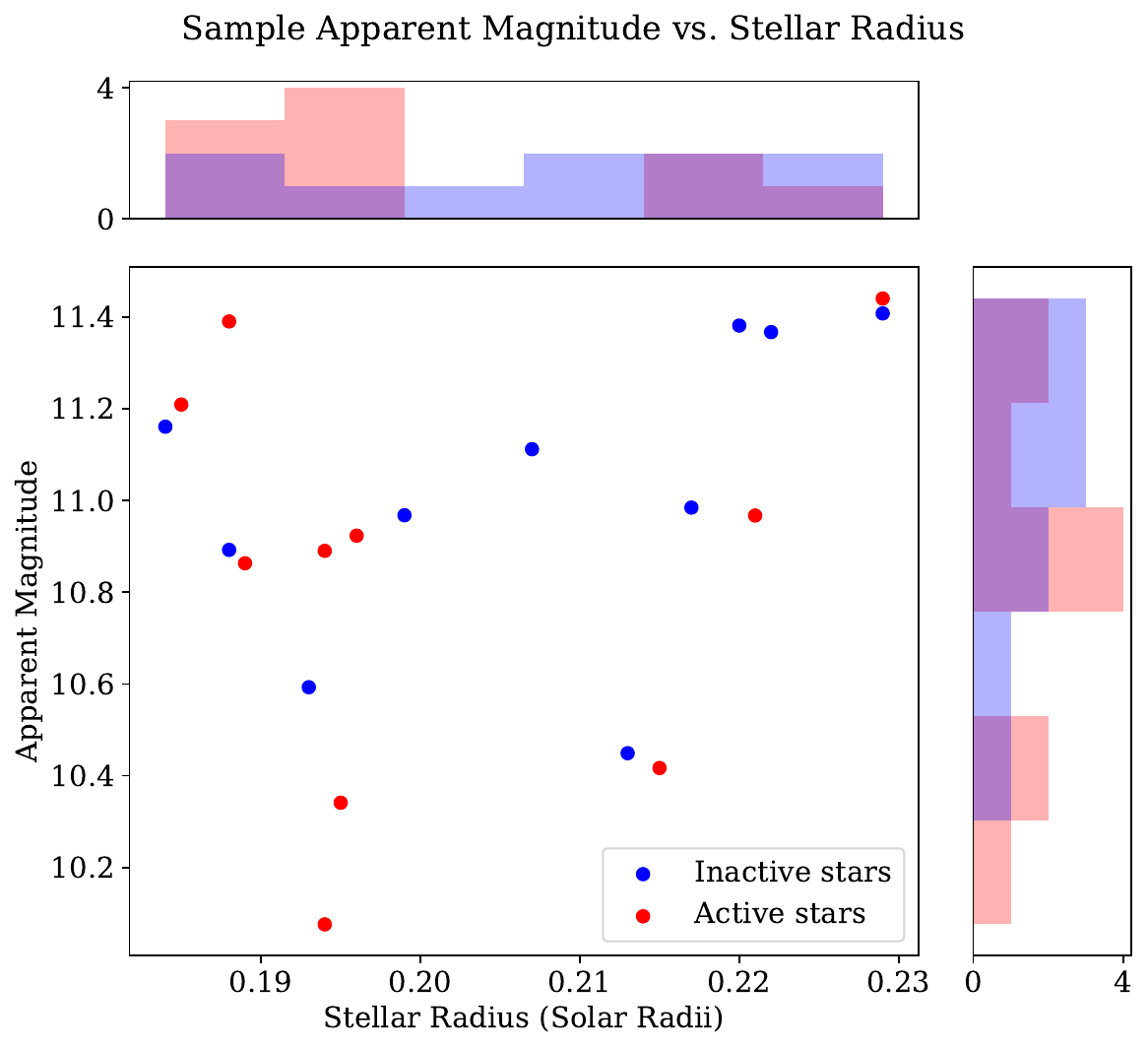}
  \caption{Our stellar sample, selected from \cite{Newton17} for extant \textit{TESS} photometry of stars of similar size and magnitude. Our sample comprises of 10 active and 10 inactive stars between $0.18$ R$_\odot$ and $0.23$ R$_\odot$ with apparent \textit{TESS} magnitudes between 10.0 and 11.5. We show histograms of these distributions above and to the side of the plot. This sample of stars are similar enough in physical properties to serve as a useful benchmark sample for this study.}
  \label{fig:histogram}
\end{figure}

 To ensure as homogeneous a sample as possible in terms of predicted transit signal-to-noise, we construct our sample to fall within a narrow range in stellar radius and apparent magnitude. We aim to characterize the effect of low-level, frequently-occurring flares on transit detection, as one of the likely contributing factors to decreased transit sensitivity for active stars. This necessitates a tradeoff in our sample criteria between flare amplitude, frequency, and predicted transit signal-to-noise. We aimed to have our marginally detectable injected transits induce the approximate same photometric offset as flares occurring with frequency similar to the ($\sim$2 hr) transit duration timescale, to ensure that  confusion between flares and transits will occur as part of our survey design. \cite{Davenport16} found that flares of 5mmag amplitude occur on average 63 times per day on the active M dwarf Proxima Centauri (roughly 1 every 20 minutes), with more energetic flares occurring less often in relationship characterized by a power law. In the \textit{TESS} magnitude range of 10-11, flares larger than this relative amplitude of 5$\times$10$^{-3}$ are detectable nearly 100\% of the time \citep{Gunter20}. We concluded that flares occurring with a frequency near a typical transit duration ($\sim$2 hours), necessarily $>$5mmag, would therefore be detectable and removable in our \textit{TESS} lightcurves if we select stars in this approximate magnitude range. Limiting ourselves to brighter targets, however, skews the sample toward earlier spectral types, for which less stars are active. In order to maintain a nominal sample size of 10 active and 10 inactive stars with a common range of stellar radius and magnitude, we identified the fiducial radius of 0.2$R_{\odot}$ as optimal for our sample: at this radius, there are (a) sufficient stars (both active and inactive) in the \cite{Newton17} sample that (b) meet the magnitude criterion for which (c) there exists at least one sector of \textit{TESS} 2 minute cadence photometry. In Table \ref{tab:sample} we list the 20 targets we decided upon: all of the stellar radii fall within the range of 0.18 to 0.23 $R_{\odot}$ ($0.05R_{\odot}$ is typical uncertainty on stellar radius, per \citealt{Berger23}). These stars also all have similar brightness, with \textit{TESS} magnitudes between 10 and 11.5. Though some stars in our sample were observed during more than one \textit{TESS} sector, we select only one sector for each star. While there exist gaps within the lightcurves dependent upon the \textit{TESS} sector in which the star was observed, we ensure that the duty cycle varies by no more than 10\% across the sample. In Figure \ref{fig:histogram} we show the \textit{TESS} magnitude and stellar radius of our resulting sample of 20 stars.

\subsection{Detrending of \textit{TESS} Lightcurves}
\label{sec:detrending}

For each of our 20 stars, we detrend the \textit{TESS} lightcurves to remove readily identifiable signatures of stellar activity.  We employ a Gaussian process model for stellar variability, using the \texttt{exoplanet} \citep{exoplanet:exoplanet} Python package. We utilize the Lomb-Scargle estimator with a minimum period of 0.1 days, a maximum period of 30.0 days, and 50 samples per peak. We modeled this detrending (into the the functional form for the Gaussian process kernel) on \cite{Ment23}, who detrended 363 \textit{TESS} lightcurves of nearly M dwarfs with a similar treatment.  

\begin{figure*}
  {\begin{tabular}{@{}cc@{}}
    \includegraphics[width=0.47\textwidth]{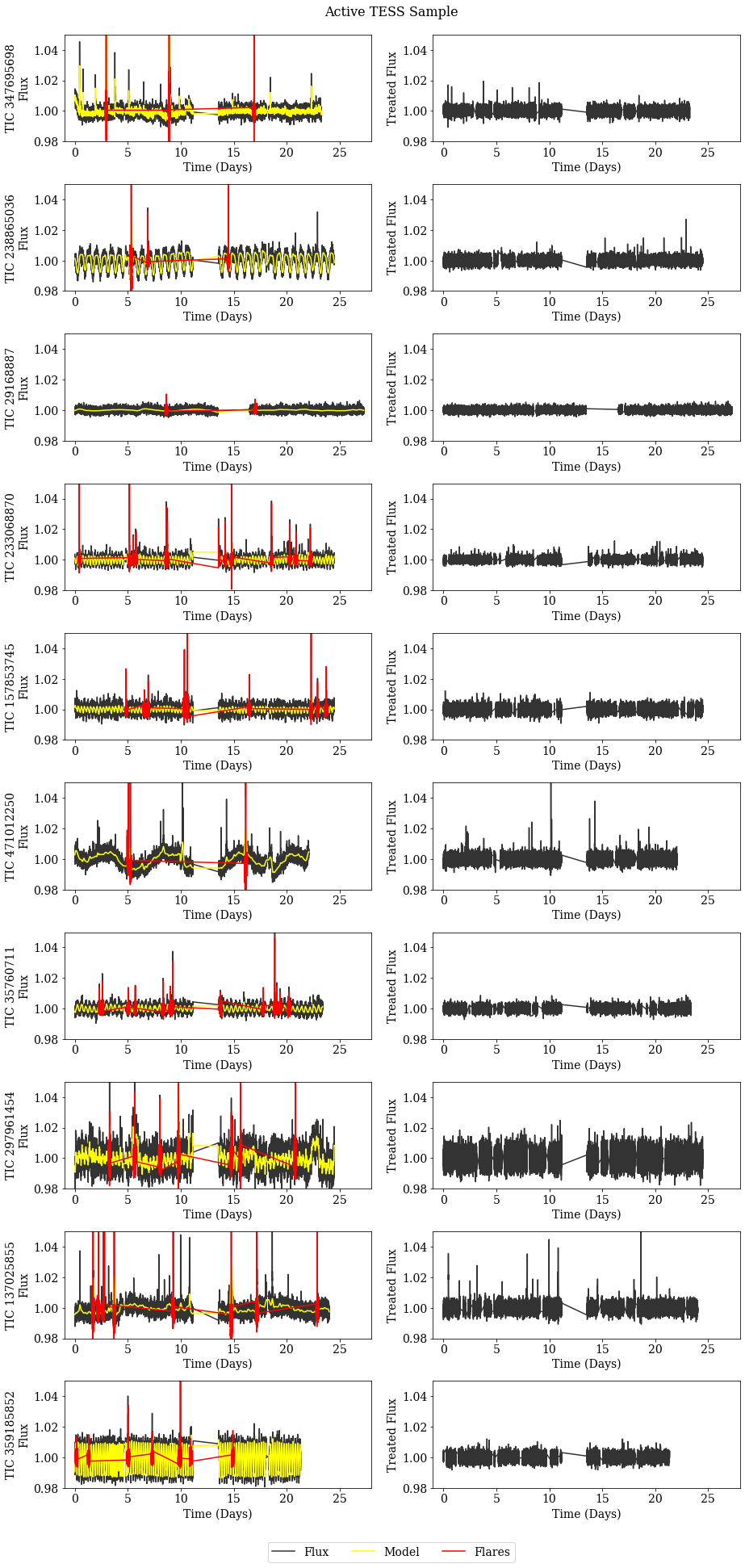}&
    \includegraphics[width=0.47\textwidth]{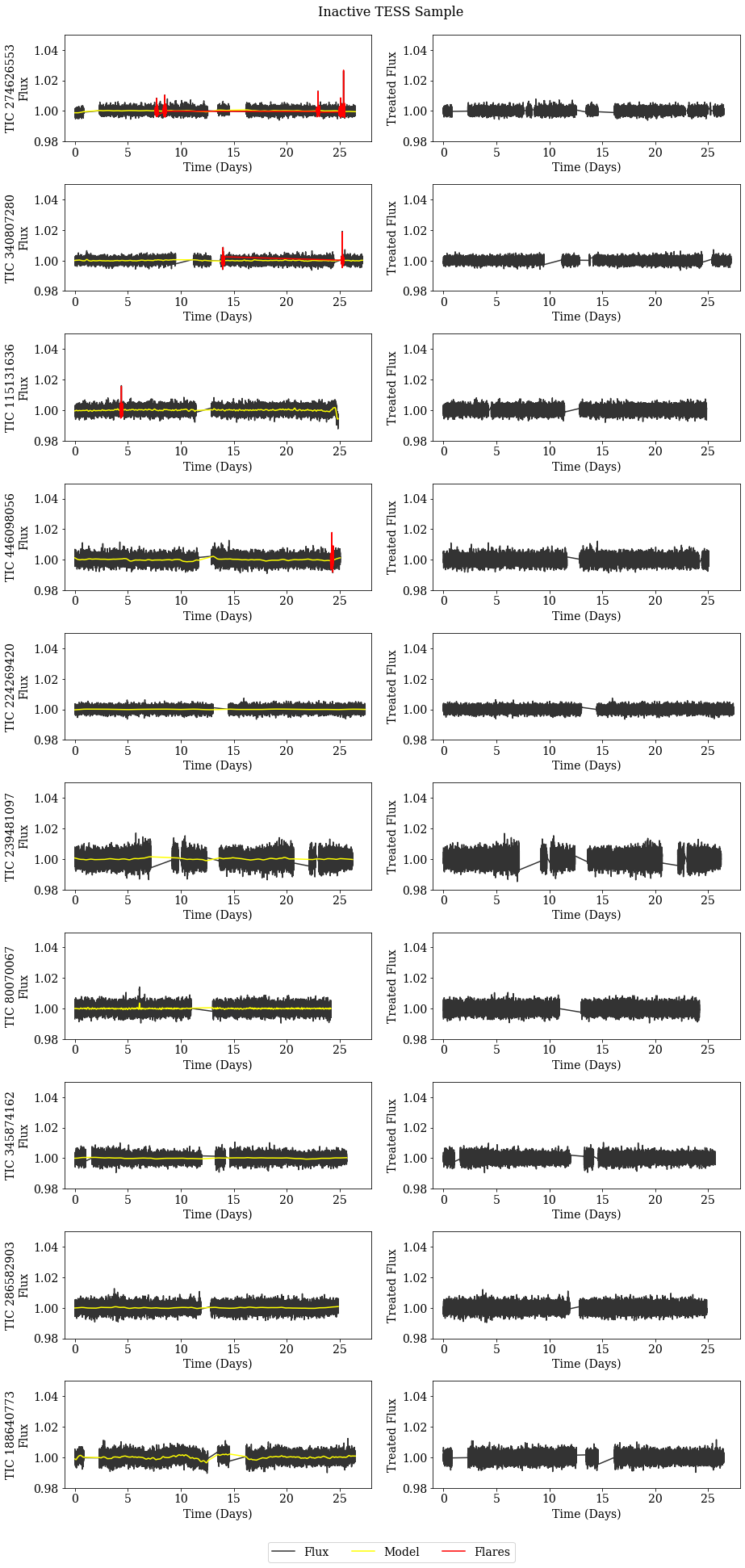}

  \end{tabular}}
    \caption{\textit{TESS} lightcurves of all twenty stars in our sample. Panel (a) at left shows the active sample and panel (b) at right shows the inactive sample. The left side of each panel shows the original non-detrended SPOC \citep{Jenkins16} photometry in black. We have overplotted the Gaussian Process stellar rotation model in yellow, and identified flare events in red. The right side of each panel displays the treated lightcurve, normalized by the rotation model and with flares (and points within 3 hours of flare data points) excised.}
  \label{fig:Active-Inactive-samples}
\end{figure*}

Upon generating the rotation model, we normalize the original lightcurve (with transits injected, see Section \ref{sec:injection}) by dividing it by the model. We identify and remove flares by identifying individual flux measurements more than 5$\sigma$ above the mean, before clipping all points within a 3-hour window of these identified points.   We apply this treatment uniformly to both active and inactive stars, ensuring  consistency in detrending across the sample. In Figure \ref{fig:Active-Inactive-samples} we show the non-detrended SPOC photometry from one sector for each of our 20 sample stars, as well as the resulting detrended photometry.  


\subsection{Injection and Recovery}
\label{sec:injection}
To account for the possibility of transit suppression from the process of detrending, we inject transits into the lightcurve prior to the detrending stage. We employ the \textit{TESS}-SPOC lightcurves \citep{Jenkins16}, and draw the stellar parameters of mass and radius from \cite{Newton17}, with a standard deviation equal to that reported in these works. 

In order to construct a sample of transiting planet parameters, we draw values in both period and radius using the log uniform distribution as \cite{Ballard19}. This choice will also enable the ultimate ease of comparison between our empirical maps we generate here and the average theoretical M dwarf sensitivity maps in that paper. We employ an inset of period parameter space from 0.8 to 5.7 days, and radius from 0.28 to 1.44 $R_{\oplus}$. We selected the center of this range in order to sample planetary signals with an approximate signal-to-noise ratio of 7, where we expected transit sensitivity to meaningfully vary (see e.g. \citealt{Christiansen16}). For each lightcurve, we inject a transiting planet with each possible combination of period and radius ten times, each time shifting the transit phase, $t_{0}$, of the planet by 3 hours per iteration in order to be sure to fully sample the effect of stellar variability on each individual star. Additionally, we assume the orbital eccentricity of all injected planets to be zero. We generate the model transit lightcurves using the \texttt{batman} \citep{Kreidberg15} Python package (normalized to unity), and inject this lightcurve into the simulated \textit{TESS} lightcurve by multiplication, completing the injection step.

Following the injection, we detrend the lightcurve, as detailed previously in Section \ref{sec:detrending}. Once the rotation model is divided out and flares are removed, we utilize the \texttt{TransitLeastSquares} \citep{Hippke19} Python package, generating a periodogram and estimating the most likely period for the transiting exoplanet. We define a transit as ``detected" if the False Alarm Probability (FAP) value is less than 0.001, and the estimated period is within 1\% of the true injected period. In this sense, we do not consider aliases of the injected period to constitute ``detection."  

\begin{figure}[h]
  \centering
    \includegraphics[width=1\linewidth]{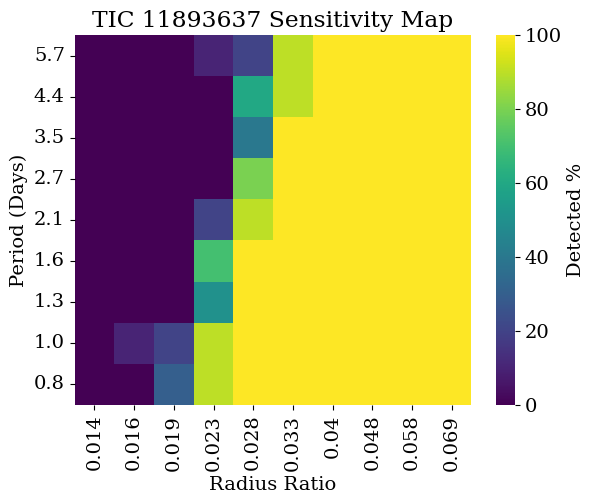}
    \caption{Sensitivity map for TIC 11893637, using the methodology described in Section \ref{sec:Methods}. While this lightcurve was ultimately not included in our final analysis due to its short duty cycle, we show a comparison to \cite{Ment23} to show consistency in completeness across methodologies. Axes drawn from \cite{Ment23}.}
  \label{fig:ment_v_ballard}
\end{figure}

To benchmark our methods against existing similar studies in the literature, we show in Figure \ref{fig:ment_v_ballard} our sensitivity map to TIC 11893637. This star is not included in our sample, but is one of the 363 M dwarfs studied by \cite{Ment23}, who also characterized the sensitivity to transits in the \textit{TESS} lightcurve. We have inverted the axes and show radius ratio, rather than planetary radius, in order to allow for direct comparison. We note broad consistency in the detection contours, though our resolution in orbital period and radius space differs. 

\section{Analysis}
\label{sec:analysis}
We apply the methodology in Section \ref{sec:Methods} to our sample of 10 active and 10 inactive M dwarf stars. In Section \ref{sec:sensitivity_compare}, we compare the sensitivity maps generated using the injection-and-recovery procedure for active and inactive stars. In Section \ref{sec:occurrence_implications}, we discuss the relevance of our findings to the detectability of transiting exoplanets orbiting active stars.

\subsection{Sensitivity in active versus inactive stars}
\label{sec:sensitivity_compare}
\begin{figure}[h]
  \centering

  \begin{subfigure}
    \centering
    \includegraphics[width=1\linewidth]{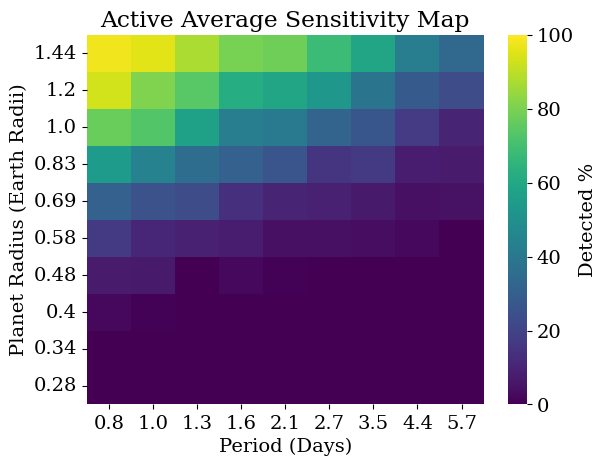}

    \label{subfig:active_heatmap}
  \end{subfigure}

  \begin{subfigure}
    \centering
    \includegraphics[width=1\linewidth]{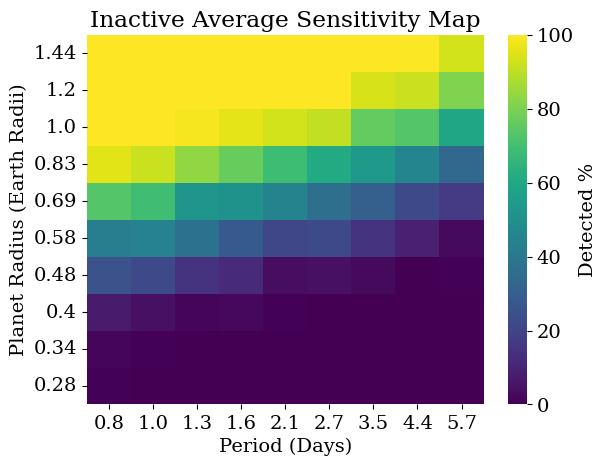}

    \label{subfig:inactive_heatmap}
  \end{subfigure}

  \caption{Detectability sensitivity map of exoplanets around active and inactive stars. These maps represent the average detectability, where ``detection" constitutes a FAP$<$0.001, and a recovery of the true period within 1\%. Detectability ranges from 0-100\%. Top: Transiting planet detectability for our active star population. We find that planet recovery decreases significantly below Earth-radii planets at periods longer than 1.6 days. Bottom: Same as the top panel but for our inactive stellar sample. We find broadly increased recovery of injected planet signals, reaching down to sub-Earth radii and periods as large as 6.0 days. We find that imperfect correction of photometric modulation can significantly degrade the ability to detect transiting planet signals.}
  \label{fig:avg_heatmaps}
\end{figure}

In Figure \ref{fig:avg_heatmaps}, we show the average sensitivity to transiting planets for each sub-sample. We constructed both the average ``active" and ``inactive" sensitivity maps from the 10individual stellar sensitivity maps in each subsample. As each star  was tested with the same number of planetary injections, each star receives uniform weight in the final average. These maps show the fraction of detected injected exoplanets in the pipeline, ranging from 0\% to 100\%. As expected, the most detactable are those with large radii and short orbital periods. However, the difference in detectability between active and active stars is apparent toward smaller radii and longer orbital periods. 

  We quantify the average extent to which transit sensitivity is degraded, by comparing common detection contours between the averaged active and inactive sensitivity maps. The departure from Gaussian noise in transit photometry typically involves characterizing the off-diagonal component of the covariance matrix, either by calculating it directly (e.g. \citealt{Pont06}, \citealt{Jenkins02} \citealt{Gillon09}) or parameterizing its behavior (e.g. \citealt{Carter09}, \citealt{Aigrain15}, or see review by \citealt{Aigrain23}). Here, as we are characterizing an average effect across stars on potentially a variety of timescales, we elect to estimate the factor $\beta$ by which the photometric uncertainty is inflated by correlated noise, per \cite{Pont06}:

  \begin{equation}
      \beta=\sqrt{1+\left(\frac{\sigma_{\textrm{r}}}{\sigma_{\textrm{w}}}\right)^{2}},
  \end{equation}

where $\sigma_{\textrm{r}}$ corresponds to the ``red" (off-diagonal) contribution to the noise budget and $\sigma_{\textrm{w}}$ corresponds to the ``white" (diagonal) contribution. For noise with no red component, $\beta=1$. Simplifying the signal-to-noise of a transit signal as a boxcar with depth $\delta$, effective uncertainty per observation $\beta\sigma$, number of observations per transit duration $N_{\textrm{obs}}$, and number of transits $N_{\textrm{transits}}$, the signal-to-noise can be expressed by

\begin{equation}
    \textrm{SNR}=\frac{\delta}{\beta\sigma}\sqrt{N_{\textrm{obs}}}\sqrt{N_{\textrm{transits}}}.
\end{equation}

If we assume a common \textit{effective} signal-to-noise ratio corresponds the same detection probability, we can quantify the extent to which $\beta$ for active stars is higher, causing a resulting degradation in signal-to-noise. Per \cite{Christiansen16} from transit searches in the \textit{Kepler} photometric pipeline, the 50\% detection likelihood corresponds to a typical effective signal-to-noise ratio of 8.  Staking our calculation to contours of common detection probability (for values between 0 and 100\%), we find that the photometric uncertainty is effectively inflated on average in stars by $\beta=$1.5--1.6. This is consistent with the amplitude of the Gaussian process kernel $\alpha$ fit to Proxima Centauri photometry by \cite{Kipping17} of 1.4--2.3.  

\begin{figure}[h]
  \centering
  \includegraphics[width=1\linewidth]{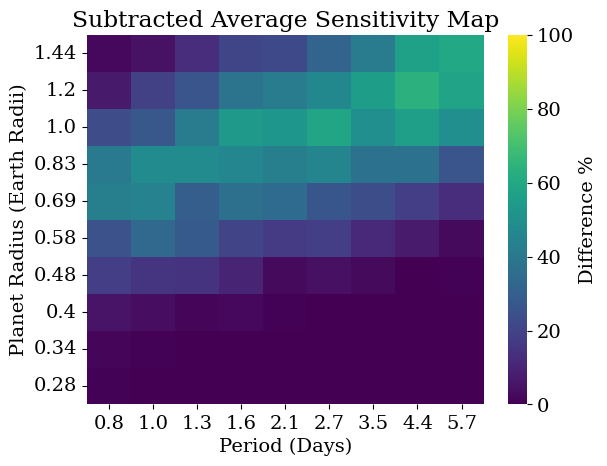}
  \caption{Inactive minus active heatmap. This map is obtained by subtracting the heatmaps displayed in Figure \ref{fig:avg_heatmaps}. We find that the inactive M dwarfs show a greater sensitivity to transiting planets in a belt spanning 0.7 R$_\oplus$ at 0.8 day periods up to 1.2 R$_\oplus$ at 6.0 day periods. }
  \label{fig:subtracted}
\end{figure}

We expect the difference between the averaged sensitivity maps to highlight the region associated with marginal detection sensitivity: the maps will similarly reflect 0\% transit probability at very long orbital periods and small radii, and 100\% transit probability at short orbital periods and large radii. In Figure \ref{fig:subtracted} we see that the difference between the detection probability of inactive versus active stars peaks at 60\% (e.g. a planet with 80\% recovery probability orbiting an inactive star would have a 20\% recovery probability when orbiting an active star). 

\subsection{Relevance for M dwarf planet occurrence}
\label{sec:occurrence_implications}
The subtracted heat map shown in Figure \ref{fig:subtracted} is the result of subtracting the active heatmap from the inactive heatmap, revealing the region where there is a discrepancy between the detectability rates between active and inactive stars. This region is significant because there are known exoplanets in this region, such as, for example, the TRAPPIST-1 planets \citep{Gillon16}. This means that despite this being a crucial region to constrain detectability due to planetary abundance, our ability to detect exoplanets around active stars in this region is depreciated. Calculation of occurrence rates-- such as in \cite{Dressing17} or meaningfully constraining models of planet formation (which require an accurate census of planetary systems)-- therefore depend on accurately correcting for this population of planets that will be ``missed" in current surveys.  

\begin{figure}[!ht]
    \centering
    \includegraphics[width=1\linewidth]{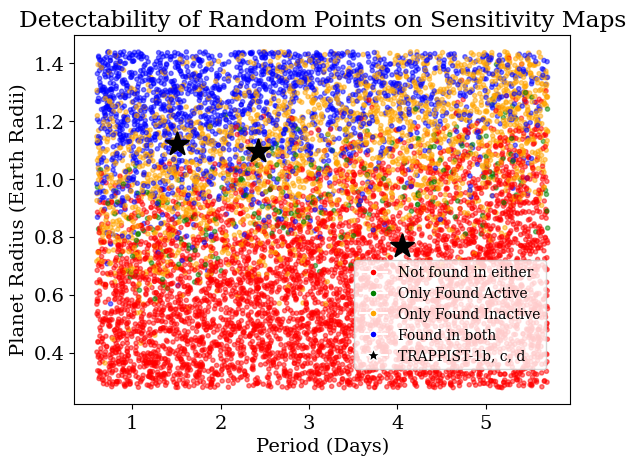}
    
    \caption{Finally, blue planets are detected around both. The orbital periods and radii of TRAPPIST-1b, c, and d \citep{Gillon16} are overplotted as star symbols. We find that some TRAPPIST-1 planets are not detectable under a \textit{TESS}-like survey and therefore existing \textit{TESS} objects of interest may in fact be a TRAPPIST-1 like system that remains unexplored or fully understood. As TRAPPIST-1 is itself an active star, this further degrades the probability of detecting a similar system blindly in TESS data.}
    \label{fig:sub-heatmaps}
\end{figure}

In Figure \ref{fig:sub-heatmaps}, we show the result of random injection-and-recovery of planets into the detectability metric from the sensitivity maps from Figure \ref{fig:avg_heatmaps}. This exercise shows the findings of Figures \ref{fig:sub-heatmaps} and \ref{fig:subtracted} in another way, where the reader can more readily see the inherent uncertainty in our sensitivity maps. We generate this plot by  selecting random points in the orbital period/radius parameter surveyed by the sensitivity maps, and then selecting a random number between 1-100, and then determining if it falls within the probability that a planet would have been detected around that type of star. If it is within the probability, it was marked as "detected" around that type of star. Our recovery fractions, as expected, deviate most strongly between ``active" and ``inactive" in the parameter space where the subtracted average maps shows the largest deviation. 

As expected, if a planet is large with a small orbital period, it is detected around both active and inactive stars. On the other extreme, small planets with large orbital periods are not detected around active nor inactive stars. The region of interest is where planets were detected by only an active or (more likely) an inactive star. Most often, stars are only detected by the inactive star in this region. As shown in \ref{fig:sub-heatmaps}, the three inner-most planets orbiting TRAPPIST-1 \citep{Gillon17} fall within this region. We emphasize here that TRAPPIST-1 itself is an ``active" host star which also is at a dimmer magnitude than the active and inactive stars surveyed here: we include them to contextualize the difficulty of identifying planets of this size and orbital period in \textit{TESS} lightcurves even of bright (10th and 11th magnitude) stars.  

\section{Conclusion}
\label{sec:conclusion}
We have investigated the detectability of transiting exoplanets around M dwarfs, with a focus upon the effects of stellar activity. By identifying a sample of common stellar radius and apparent magnitude, that differs only in ``activity" status, we can directly assess the extent to which stellar activity (which occurs over a wide range of timescales and amplitudes) affects the detection of transiting planets in practice. We crafted our sample in order to allow for confusion between (a) flares occurring with frequency near the typical $\sim$2 hour transit duration and (b) marginally detectable transiting planets (that is, the two have similar predicted amplitude). Of course, this finding is limited in scope, as flaring on all timescales will affect transit detection, as well the detrending necessarily to remove stellar rotation. As predicted, our sensitivity to transiting planets is degraded around active stars. We quantify this affect by estimating $\beta$, the inflation of the photometric uncertainty in active stellar lightcurves due to the presence of correlated noise. While we have limited our study here to stars with radius $\sim$0.2$R_{\odot}$ and \textit{TESS} magnitudes between 10-11.5, the estimation of $\beta$ to be between 1.5 and 1.6 may be useful in a broader context as a normalization in the sensitivity to transits around active versus inactive M dwarfs.   

We find that the types of planets that are detected around inactive but not inactive stars fall in an important parameter space where terrestrial exoplanets are known to occur. Particularly, we find that a realistic noise budget for active stars makes the detection of additional TRAPPIST-1 like systems challenging in current \textit{TESS} data even of bright stars. The use of TESS data for the calculation of occurrence rate, particularly at the low mass end of the stellar main sequence where active lifetimes are long and stellar variability is high, requires a more thorough understanding of how variability manifests in lightcurves. Future work quantifying this effect is warranted as well as future work exploring other methods for mitigating the effect of stellar activity on transit detection. The habitable zone of these stars falls in the regions of interest explored in this work, and so developing a census and target list of potentially habitable rocky planets for future missions will heavily depend on our ability to mitigate and understand stellar activity in these transit lightcurves. 

\section{Acknowledgements}
\label{sec:Acknowledgements}

We thank  Quadry Chance, Sheila Sagear, Christopher Lam, Natalia Guerrero, William Schap III, and Kristo Ment for useful discussions that have helped inform and enrich this work. This material is based on work supported in part by the University of Florida Undergraduate Scholars Program and the William Oegerle Scholarship in Physics and Astronomy. 

We acknowledge that for thousands of years the area now comprising the state of Florida has been, and continues to be, home to many Native Nations. We further recognize that the main campus of the University of Florida is located on the ancestral territory of the Potano and of the Seminole peoples. The Potano, of Timucua affiliation, lived here in the Alachua region from before European arrival until the destruction of their towns in the early 1700s. The Seminole, also known as the Alachua Seminole, established towns here shortly after but were forced from the land as a result of a series of wars with the United States known as the Seminole Wars. We, the authors, acknowledge our obligation to honor the past, present, and future Native residents and cultures of Florida.

This research made use of \textsf{exoplanet} \citep{exoplanet:joss,
exoplanet:zenodo} and its dependencies \citep{exoplanet:agol20,
exoplanet:arviz, exoplanet:astropy13, exoplanet:astropy18, exoplanet:kipping13,
exoplanet:luger18, exoplanet:pymc3, exoplanet:theano}.

\software{exoplanet 
\citep{exoplanet:exoplanet},
batman \citep{Kreidberg15} , TransitLeastSquares \citep{Hippke19}}

\bibliography{main.bib}

\begin{thebibliography}{70}
\expandafter\ifx\csname natexlab\endcsname\relax\def\natexlab#1{#1}\fi

\bibitem[{{Agol} {et~al.}(2020){Agol}, {Luger}, \&
  {Foreman-Mackey}}]{exoplanet:agol20}
{Agol}, E., {Luger}, R., \& {Foreman-Mackey}, D. 2020, \aj, 159, 123

\bibitem[{{Aigrain} \& {Foreman-Mackey}(2023)}]{Aigrain23}
{Aigrain}, S., \& {Foreman-Mackey}, D. 2023, \araa, 61, 329

\bibitem[{{Aigrain} {et~al.}(2015){Aigrain}, {Llama}, {Ceillier}, {Chagas},
  {Davenport}, {Garc{\'\i}a}, {Hay}, {Lanza}, {McQuillan}, {Mazeh}, {de
  Medeiros}, {Nielsen}, \& {Reinhold}}]{Aigrain15}
{Aigrain}, S., {et~al.} 2015, \mnras, 450, 3211

\bibitem[{{Astropy Collaboration} {et~al.}(2018){Astropy Collaboration},
  {Price-Whelan}, {Sip{\H o}cz}, {G{\"u}nther}, {Lim}, {Crawford}, {Conseil},
  {Shupe}, {Craig}, {Dencheva}, {Ginsburg}, {VanderPlas}, {Bradley},
  {P{\'e}rez-Su{\'a}rez}, {de Val-Borro}, {Aldcroft}, {Cruz}, {Robitaille},
  {Tollerud}, {Ardelean}, {Babej}, {Bach}, {Bachetti}, {Bakanov}, {Bamford},
  {Barentsen}, {Barmby}, {Baumbach}, {Berry}, {Biscani}, {Boquien}, {Bostroem},
  {Bouma}, {Brammer}, {Bray}, {Breytenbach}, {Buddelmeijer}, {Burke},
  {Calderone}, {Cano Rodr{\'{\i}}guez}, {Cara}, {Cardoso}, {Cheedella},
  {Copin}, {Corrales}, {Crichton}, {D'Avella}, {Deil}, {Depagne}, {Dietrich},
  {Donath}, {Droettboom}, {Earl}, {Erben}, {Fabbro}, {Ferreira}, {Finethy},
  {Fox}, {Garrison}, {Gibbons}, {Goldstein}, {Gommers}, {Greco}, {Greenfield},
  {Groener}, {Grollier}, {Hagen}, {Hirst}, {Homeier}, {Horton}, {Hosseinzadeh},
  {Hu}, {Hunkeler}, {Ivezi{\'c}}, {Jain}, {Jenness}, {Kanarek}, {Kendrew},
  {Kern}, {Kerzendorf}, {Khvalko}, {King}, {Kirkby}, {Kulkarni}, {Kumar},
  {Lee}, {Lenz}, {Littlefair}, {Ma}, {Macleod}, {Mastropietro}, {McCully},
  {Montagnac}, {Morris}, {Mueller}, {Mumford}, {Muna}, {Murphy}, {Nelson},
  {Nguyen}, {Ninan}, {N{\"o}the}, {Ogaz}, {Oh}, {Parejko}, {Parley}, {Pascual},
  {Patil}, {Patil}, {Plunkett}, {Prochaska}, {Rastogi}, {Reddy Janga},
  {Sabater}, {Sakurikar}, {Seifert}, {Sherbert}, {Sherwood-Taylor}, {Shih},
  {Sick}, {Silbiger}, {Singanamalla}, {Singer}, {Sladen}, {Sooley},
  {Sornarajah}, {Streicher}, {Teuben}, {Thomas}, {Tremblay}, {Turner},
  {Terr{\'o}n}, {van Kerkwijk}, {de la Vega}, {Watkins}, {Weaver}, {Whitmore},
  {Woillez}, {Zabalza}, \& {Astropy Contributors}}]{exoplanet:astropy18}
{Astropy Collaboration} {et~al.} 2018, \aj, 156, 123

\bibitem[{{Astropy Collaboration} {et~al.}(2013){Astropy Collaboration},
  {Robitaille}, {Tollerud}, {Greenfield}, {Droettboom}, {Bray}, {Aldcroft},
  {Davis}, {Ginsburg}, {Price-Whelan}, {Kerzendorf}, {Conley}, {Crighton},
  {Barbary}, {Muna}, {Ferguson}, {Grollier}, {Parikh}, {Nair}, {Unther},
  {Deil}, {Woillez}, {Conseil}, {Kramer}, {Turner}, {Singer}, {Fox}, {Weaver},
  {Zabalza}, {Edwards}, {Azalee Bostroem}, {Burke}, {Casey}, {Crawford},
  {Dencheva}, {Ely}, {Jenness}, {Labrie}, {Lim}, {Pierfederici}, {Pontzen},
  {Ptak}, {Refsdal}, {Servillat}, \& {Streicher}}]{exoplanet:astropy13}
---. 2013, \aap, 558, A33

\bibitem[{{Ballard}(2019)}]{Ballard19}
{Ballard}, S. 2019, \aj, 157, 113

\bibitem[{Barnes {et~al.}(2004)Barnes, James, \& Collier~Cameron}]{Barnes2004}
Barnes, J.~R., James, D.~J., \& Collier~Cameron, A. 2004, Monthly Notices of
  the Royal Astronomical Society, 352, 589

\bibitem[{{Basri} {et~al.}(2013){Basri}, {Walkowicz}, \& {Reiners}}]{Basri13}
{Basri}, G., {Walkowicz}, L.~M., \& {Reiners}, A. 2013, \apj, 769, 37

\bibitem[{{Bastien} {et~al.}(2013){Bastien}, {Stassun}, {Basri}, \&
  {Pepper}}]{Bastien13}
{Bastien}, F.~A., {Stassun}, K.~G., {Basri}, G., \& {Pepper}, J. 2013, \nat,
  500, 427

\bibitem[{{Bastien} {et~al.}(2016){Bastien}, {Stassun}, {Basri}, \&
  {Pepper}}]{Bastien16}
---. 2016, \apj, 818, 43

\bibitem[{{Bell} {et~al.}(2012){Bell}, {Hilton}, {Davenport}, {Hawley}, {West},
  \& {Rogel}}]{Bell12}
{Bell}, K.~J., {Hilton}, E.~J., {Davenport}, J. R.~A., {Hawley}, S.~L., {West},
  A.~A., \& {Rogel}, A.~B. 2012, \pasp, 124, 14

\bibitem[{{Berger} {et~al.}(2023){Berger}, {Schlieder}, \& {Huber}}]{Berger23}
{Berger}, T.~A., {Schlieder}, J.~E., \& {Huber}, D. 2023, arXiv e-prints,
  arXiv:2301.11338

\bibitem[{{Boisse} {et~al.}(2012){Boisse}, {Bonfils}, \& {Santos}}]{Boisse12}
{Boisse}, I., {Bonfils}, X., \& {Santos}, N.~C. 2012, \aap, 545, A109

\bibitem[{{Carter} \& {Winn}(2009)}]{Carter09}
{Carter}, J.~A., \& {Winn}, J.~N. 2009, \apj, 704, 51

\bibitem[{{Cauley} {et~al.}(2017){Cauley}, {Redfield}, \&
  {Jensen}}]{Cauley2017}
{Cauley}, P.~W., {Redfield}, S., \& {Jensen}, A.~G. 2017, \aj, 153, 217

\bibitem[{{Christiansen} {et~al.}(2016){Christiansen}, {Clarke}, {Burke},
  {Jenkins}, {Bryson}, {Coughlin}, {Mullally}, {Thompson}, {Twicken},
  {Batalha}, {Haas}, {Catanzarite}, {Campbell}, {Kamal Uddin}, {Zamudio},
  {Smith}, \& {Henze}}]{Christiansen16}
{Christiansen}, J.~L., {et~al.} 2016, \apj, 828, 99

\bibitem[{{Davenport}(2016)}]{Davenport16}
{Davenport}, J. R.~A. 2016, \apj, 829, 23

\bibitem[{Delfosse {et~al.}(1998)Delfosse, Forveille, Perrier, \&
  Mayor}]{Delfosse98}
Delfosse, X., Forveille, T., Perrier, C., \& Mayor, M. 1998, Astronomy and
  Astrophysics, 331, 581

\bibitem[{{Dressing} \& {Charbonneau}(2015)}]{Dressing15}
{Dressing}, C.~D., \& {Charbonneau}, D. 2015, \apj, 807, 45

\bibitem[{{Dressing} {et~al.}(2017){Dressing}, {Vanderburg}, {Schlieder},
  {Crossfield}, {Knutson}, {Newton}, {Ciardi}, {Fulton}, {Gonzales}, {Howard},
  {Isaacson}, {Livingston}, {Petigura}, {Sinukoff}, {Everett}, {Horch}, \&
  {Howell}}]{Dressing17}
{Dressing}, C.~D., {et~al.} 2017, ArXiv e-prints

\bibitem[{{Dumusque} {et~al.}(2014){Dumusque}, {Boisse}, \&
  {Santos}}]{Dumusque14}
{Dumusque}, X., {Boisse}, I., \& {Santos}, N.~C. 2014, \apj, 796, 132

\bibitem[{{Foreman-Mackey} {et~al.}(2021){Foreman-Mackey}, {Luger}, {Agol},
  {Barclay}, {Bouma}, {Brandt}, {Czekala}, {David}, {Dong}, {Gilbert},
  {Gordon}, {Hedges}, {Hey}, {Morris}, {Price-Whelan}, \&
  {Savel}}]{exoplanet:joss}
{Foreman-Mackey}, D., {et~al.} 2021, arXiv e-prints, arXiv:2105.01994

\bibitem[{Foreman-Mackey {et~al.}(2021{\natexlab{a}})Foreman-Mackey, Savel,
  Luger, Agol, Czekala, Price-Whelan, Hedges, Gilbert, Bouma, Brandt, \&
  Barclay}]{exoplanet:zenodo}
Foreman-Mackey, D., {et~al.} 2021{\natexlab{a}}, exoplanet-dev/exoplanet v0.5.1

\bibitem[{Foreman-Mackey {et~al.}(2021{\natexlab{b}})Foreman-Mackey, Savel,
  Luger, Czekala, Agol, Price-Whelan, Hedges, Gilbert, Barclay, Bouma, \&
  Brandt}]{exoplanet:exoplanet}
---. 2021{\natexlab{b}}, exoplanet-dev/exoplanet v0.4.5

\bibitem[{{Giampapa} \& {Liebert}(1986)}]{Giampapa86}
{Giampapa}, M.~S., \& {Liebert}, J. 1986, \apj, 305, 784

\bibitem[{{Gillon} {et~al.}(2016){Gillon}, {Jehin}, {Lederer}, {Delrez}, {de
  Wit}, {Burdanov}, {Van Grootel}, {Burgasser}, {Triaud}, {Opitom}, {Demory},
  {Sahu}, {Bardalez Gagliuffi}, {Magain}, \& {Queloz}}]{Gillon16}
{Gillon}, M., {et~al.} 2016, \nat, 533, 221

\bibitem[{{Gillon} {et~al.}(2009){Gillon}, {Smalley}, {Hebb}, {Anderson},
  {Triaud}, {Hellier}, {Maxted}, {Queloz}, \& {Wilson}}]{Gillon09}
---. 2009, \aap, 496, 259

\bibitem[{{Gillon} {et~al.}(2017){Gillon}, {Triaud}, {Demory}, {Jehin}, {Agol},
  {Deck}, {Lederer}, {de Wit}, {Burdanov}, {Ingalls}, {Bolmont}, {Leconte},
  {Raymond}, {Selsis}, {Turbet}, {Barkaoui}, {Burgasser}, {Burleigh}, {Carey},
  {Chaushev}, {Copperwheat}, {Delrez}, {Fernandes}, {Holdsworth}, {Kotze}, {Van
  Grootel}, {Almleaky}, {Benkhaldoun}, {Magain}, \& {Queloz}}]{Gillon17}
---. 2017, \nat, 542, 456

\bibitem[{{Guerrero} {et~al.}(2021){Guerrero}, {Seager}, {Huang}, {Vanderburg},
  {Garcia Soto}, {Mireles}, {Hesse}, {Fong}, {Glidden}, {Shporer}, {Latham},
  {Collins}, {Quinn}, {Burt}, {Dragomir}, {Crossfield}, {Vanderspek},
  {Fausnaugh}, {Burke}, {Ricker}, {Daylan}, {Essack}, {G{\"u}nther}, {Osborn},
  {Pepper}, {Rowden}, {Sha}, {Villanueva}, {Yahalomi}, {Yu}, {Ballard},
  {Batalha}, {Berardo}, {Chontos}, {Dittmann}, {Esquerdo}, {Mikal-Evans},
  {Jayaraman}, {Krishnamurthy}, {Louie}, {Mehrle}, {Niraula}, {Rackham},
  {Rodriguez}, {Rowden}, {Sousa-Silva}, {Watanabe}, {Wong}, {Zhan},
  {Zivanovic}, {Christiansen}, {Ciardi}, {Swain}, {Lund}, {Mullally},
  {Fleming}, {Rodriguez}, {Boyd}, {Quintana}, {Barclay}, {Col{\'o}n},
  {Rinehart}, {Schlieder}, {Clampin}, {Jenkins}, {Twicken}, {Caldwell},
  {Coughlin}, {Henze}, {Lissauer}, {Morris}, {Rose}, {Smith}, {Tenenbaum},
  {Ting}, {Wohler}, {Bakos}, {Bean}, {Berta-Thompson}, {Bieryla}, {Bouma},
  {Buchhave}, {Butler}, {Charbonneau}, {Doty}, {Ge}, {Holman}, {Howard},
  {Kaltenegger}, {Kane}, {Kjeldsen}, {Kreidberg}, {Lin}, {Minsky}, {Narita},
  {Paegert}, {P{\'a}l}, {Palle}, {Sasselov}, {Spencer}, {Sozzetti}, {Stassun},
  {Torres}, {Udry}, \& {Winn}}]{Guerrero}
{Guerrero}, N.~M., {et~al.} 2021, \apjs, 254, 39

\bibitem[{{G{\"u}nther} {et~al.}(2020){G{\"u}nther}, {Zhan}, {Seager},
  {Rimmer}, {Ranjan}, {Stassun}, {Oelkers}, {Daylan}, {Newton}, {Kristiansen},
  {Olah}, {Gillen}, {Rappaport}, {Ricker}, {Vanderspek}, {Latham}, {Winn},
  {Jenkins}, {Glidden}, {Fausnaugh}, {Levine}, {Dittmann}, {Quinn},
  {Krishnamurthy}, \& {Ting}}]{Gunter20}
{G{\"u}nther}, M.~N., {et~al.} 2020, \aj, 159, 60

\bibitem[{{Hardegree-Ullman} {et~al.}(2019){Hardegree-Ullman}, {Cushing},
  {Muirhead}, \& {Christiansen}}]{Hardegree19}
{Hardegree-Ullman}, K.~K., {Cushing}, M.~C., {Muirhead}, P.~S., \&
  {Christiansen}, J.~L. 2019, \aj, 158, 75

\bibitem[{{Hawley} {et~al.}(2014){Hawley}, {Davenport}, {Kowalski},
  {Wisniewski}, {Hebb}, {Deitrick}, \& {Hilton}}]{Hawley14}
{Hawley}, S.~L., {Davenport}, J. R.~A., {Kowalski}, A.~F., {Wisniewski}, J.~P.,
  {Hebb}, L., {Deitrick}, R., \& {Hilton}, E.~J. 2014, \apj, 797, 121

\bibitem[{Hawley {et~al.}(1996)Hawley, Gizis, \& Reid}]{Hawley96}
Hawley, S.~L., Gizis, J.~E., \& Reid, I.~N. 1996, Astronomical Journal, 112,
  2799

\bibitem[{Hawley {et~al.}(1999)Hawley, Reid, Gizis, \& Byrne}]{Hawley99}
Hawley, S.~L., Reid, I.~N., Gizis, J.~E., \& Byrne, P.~B. 1999, in Solar and
  Stellar Activity: Similarities and Differences, ASP Conference Series, ed.
  C.~J. Butler \& J.~G. Doyle, Vol. 158, 63

\bibitem[{{He} {et~al.}(2020){He}, {Ford}, {Ragozzine}, \& {Carrera}}]{He20}
{He}, M.~Y., {Ford}, E.~B., {Ragozzine}, D., \& {Carrera}, D. 2020, \aj, 160,
  276

\bibitem[{{Henry} {et~al.}(2004){Henry}, {Subasavage}, {Brown}, {Beaulieu},
  {Jao}, \& {Hambly}}]{Henry04}
{Henry}, T.~J., {Subasavage}, J.~P., {Brown}, M.~A., {Beaulieu}, T.~D., {Jao},
  W.-C., \& {Hambly}, N.~C. 2004, \aj, 128, 2460

\bibitem[{{Herrero} {et~al.}(2016){Herrero}, {Ribas}, {Jordi}, {Morales},
  {Perger}, \& {Rosich}}]{Herrero16}
{Herrero}, E., {Ribas}, I., {Jordi}, C., {Morales}, J.~C., {Perger}, M., \&
  {Rosich}, A. 2016, \aap, 586, A131

\bibitem[{{Hippke} \& {Heller}(2019)}]{Hippke19}
{Hippke}, M., \& {Heller}, R. 2019, \aap, 623, A39

\bibitem[{{Irwin} {et~al.}(2011){Irwin}, {Berta}, {Burke}, {Charbonneau},
  {Nutzman}, {West}, \& {Falco}}]{Irwin11}
{Irwin}, J., {Berta}, Z.~K., {Burke}, C.~J., {Charbonneau}, D., {Nutzman}, P.,
  {West}, A.~A., \& {Falco}, E.~E. 2011, \apj, 727, 56

\bibitem[{{Jenkins} {et~al.}(2002){Jenkins}, {Caldwell}, \&
  {Borucki}}]{Jenkins02}
{Jenkins}, J.~M., {Caldwell}, D.~A., \& {Borucki}, W.~J. 2002, \apj, 564, 495

\bibitem[{{Jenkins} {et~al.}(2016){Jenkins}, {Twicken}, {McCauliff},
  {Campbell}, {Sanderfer}, {Lung}, {Mansouri-Samani}, {Girouard}, {Tenenbaum},
  {Klaus}, {Smith}, {Caldwell}, {Chacon}, {Henze}, {Heiges}, {Latham},
  {Morgan}, {Swade}, {Rinehart}, \& {Vanderspek}}]{Jenkins16}
{Jenkins}, J.~M., {et~al.} 2016, in Society of Photo-Optical Instrumentation
  Engineers (SPIE) Conference Series, Vol. 9913, Software and
  Cyberinfrastructure for Astronomy IV, ed. G.~{Chiozzi} \& J.~C. {Guzman},
  99133E

\bibitem[{{Kiman} {et~al.}(2021){Kiman}, {Faherty}, {Cruz}, {Gagn{\'e}},
  {Angus}, {Schmidt}, {Mann}, {Bardalez Gagliuffi}, \& {Rice}}]{Kiman21}
{Kiman}, R., {et~al.} 2021, \aj, 161, 277

\bibitem[{{Kipping}(2013)}]{exoplanet:kipping13}
{Kipping}, D.~M. 2013, \mnras, 435, 2152

\bibitem[{{Kipping} {et~al.}(2017){Kipping}, {Cameron}, {Hartman}, {Davenport},
  {Matthews}, {Sasselov}, {Rowe}, {Siverd}, {Chen}, {Sandford}, {Bakos},
  {Jord{\'a}n}, {Bayliss}, {Henning}, {Mancini}, {Penev}, {Csubry}, {Bhatti},
  {Da Silva Bento}, {Guenther}, {Kuschnig}, {Moffat}, {Rucinski}, \&
  {Weiss}}]{Kipping17}
{Kipping}, D.~M., {et~al.} 2017, \aj, 153, 93

\bibitem[{{Kreidberg}(2015)}]{Kreidberg15}
{Kreidberg}, L. 2015, \pasp, 127, 1161

\bibitem[{{Kruse} {et~al.}(2010){Kruse}, {Berger}, {Knapp}, {Laskar}, {Gunn},
  {Loomis}, {Lupton}, \& {Schlegel}}]{Kruse10}
{Kruse}, E.~A., {Berger}, E., {Knapp}, G.~R., {Laskar}, T., {Gunn}, J.~E.,
  {Loomis}, C.~P., {Lupton}, R.~H., \& {Schlegel}, D.~J. 2010, \apj, 722, 1352

\bibitem[{Kumar {et~al.}(2019)Kumar, Carroll, Hartikainen, \&
  Martin}]{exoplanet:arviz}
Kumar, R., Carroll, C., Hartikainen, A., \& Martin, O.~A. 2019, The Journal of
  Open Source Software

\bibitem[{{Luger} {et~al.}(2019){Luger}, {Agol}, {Foreman-Mackey}, {Fleming},
  {Lustig-Yaeger}, \& {Deitrick}}]{exoplanet:luger18}
{Luger}, R., {Agol}, E., {Foreman-Mackey}, D., {Fleming}, D.~P.,
  {Lustig-Yaeger}, J., \& {Deitrick}, R. 2019, \aj, 157, 64

\bibitem[{{Medina} {et~al.}(2022){Medina}, {Charbonneau}, {Winters}, {Irwin},
  \& {Mink}}]{Medina22}
{Medina}, A.~A., {Charbonneau}, D., {Winters}, J.~G., {Irwin}, J., \& {Mink},
  J. 2022, \apj, 928, 185

\bibitem[{{Ment} \& {Charbonneau}(2023)}]{Ment23}
{Ment}, K., \& {Charbonneau}, D. 2023, \aj, 165, 265

\bibitem[{{Mignon} {et~al.}(2023){Mignon}, {Meunier}, {Delfosse}, {Bonfils},
  {Santos}, {Forveille}, {Gaisn{\'e}}, {Astudillo-Defru}, {Lovis}, \&
  {Udry}}]{Mignon23}
{Mignon}, L., {et~al.} 2023, \aap, 675, A168

\bibitem[{{Muirhead} {et~al.}(2017){Muirhead}, {Dressing}, {Mann},
  {Rojas-Ayala}, {Lepine}, {Paegert}, {De Lee}, \& {Oelkers}}]{Muirhead17}
{Muirhead}, P.~S., {Dressing}, C., {Mann}, A.~W., {Rojas-Ayala}, B., {Lepine},
  S., {Paegert}, M., {De Lee}, N., \& {Oelkers}, R. 2017, \apj, submitted
  (arXiv:1710.00193)

\bibitem[{{Mulders} {et~al.}(2015){Mulders}, {Pascucci}, \& {Apai}}]{Mulders15}
{Mulders}, G.~D., {Pascucci}, I., \& {Apai}, D. 2015, \apj, 814, 130

\bibitem[{{Newton} {et~al.}(2017){Newton}, {Irwin}, {Charbonneau}, {Berlind},
  {Calkins}, \& {Mink}}]{Newton17}
{Newton}, E.~R., {Irwin}, J., {Charbonneau}, D., {Berlind}, P., {Calkins},
  M.~L., \& {Mink}, J. 2017, \apj, 834, 85

\bibitem[{{Newton} {et~al.}(2016){Newton}, {Irwin}, {Charbonneau},
  {Berta-Thompson}, {Dittmann}, \& {West}}]{Newton2016}
{Newton}, E.~R., {Irwin}, J., {Charbonneau}, D., {Berta-Thompson}, Z.~K.,
  {Dittmann}, J.~A., \& {West}, A.~A. 2016, \apj, 821, 93

\bibitem[{{O'Neal} {et~al.}(2004){O'Neal}, {Neff}, {Saar}, \&
  {Cuntz}}]{Oneal04}
{O'Neal}, D., {Neff}, J.~E., {Saar}, S.~H., \& {Cuntz}, M. 2004, \aj, 128, 1802

\bibitem[{{Pineda} {et~al.}(2013){Pineda}, {West}, {Bochanski}, \&
  {Burgasser}}]{Pineda13}
{Pineda}, J.~S., {West}, A.~A., {Bochanski}, J.~J., \& {Burgasser}, A.~J. 2013,
  \aj, 146, 50

\bibitem[{{Pont} {et~al.}(2006){Pont}, {Zucker}, \& {Queloz}}]{Pont06}
{Pont}, F., {Zucker}, S., \& {Queloz}, D. 2006, \mnras, 373, 231

\bibitem[{{Popinchalk} {et~al.}(2021){Popinchalk}, {Faherty}, {Kiman},
  {Gagn{\'e}}, {Curtis}, {Angus}, {Cruz}, \& {Rice}}]{Popinchalk21}
{Popinchalk}, M., {Faherty}, J.~K., {Kiman}, R., {Gagn{\'e}}, J., {Curtis},
  J.~L., {Angus}, R., {Cruz}, K.~L., \& {Rice}, E.~L. 2021, \apj, 916, 77

\bibitem[{{Reid} {et~al.}(1995){Reid}, {Hawley}, \& {Gizis}}]{Reid95}
{Reid}, I.~N., {Hawley}, S.~L., \& {Gizis}, J.~E. 1995, \aj, 110, 1838

\bibitem[{{Ricker} {et~al.}(2014){Ricker}, {Winn}, {Vanderspek}, {Latham},
  {Bakos}, {Bean}, {Berta-Thompson}, {Brown}, {Buchhave}, {Butler}, {Butler},
  {Chaplin}, {Charbonneau}, {Christensen-Dalsgaard}, {Clampin}, {Deming},
  {Doty}, {De Lee}, {Dressing}, {Dunham}, {Endl}, {Fressin}, {Ge}, {Henning},
  {Holman}, {Howard}, {Ida}, {Jenkins}, {Jernigan}, {Johnson}, {Kaltenegger},
  {Kawai}, {Kjeldsen}, {Laughlin}, {Levine}, {Lin}, {Lissauer}, {MacQueen},
  {Marcy}, {McCullough}, {Morton}, {Narita}, {Paegert}, {Palle}, {Pepe},
  {Pepper}, {Quirrenbach}, {Rinehart}, {Sasselov}, {Sato}, {Seager},
  {Sozzetti}, {Stassun}, {Sullivan}, {Szentgyorgyi}, {Torres}, {Udry}, \&
  {Villasenor}}]{Ricker14}
{Ricker}, G.~R., {et~al.} 2014, in \procspie, Vol. 9143, Space Telescopes and
  Instrumentation 2014: Optical, Infrared, and Millimeter Wave, 914320

\bibitem[{Salvatier {et~al.}(2016)Salvatier, Wiecki, \&
  Fonnesbeck}]{exoplanet:pymc3}
Salvatier, J., Wiecki, T.~V., \& Fonnesbeck, C. 2016, PeerJ Computer Science,
  2, e55

\bibitem[{{Soderblom} {et~al.}(1991){Soderblom}, {Duncan}, \&
  {Johnson}}]{Soderblom91}
{Soderblom}, D.~R., {Duncan}, D.~K., \& {Johnson}, D. R.~H. 1991, \apj, 375,
  722

\bibitem[{Soto {et~al.}(2023)Soto, Newton, Douglas, Burrows, \&
  Kesseli}]{GarciaSoto2023}
Soto, A.~G., Newton, E.~R., Douglas, S.~T., Burrows, A., \& Kesseli, A.~Y.
  2023, The Astronomical Journal, 165, 192, published 2023 April 5

\bibitem[{{Stassun} {et~al.}(2019){Stassun}, {Oelkers}, {Paegert}, {Torres},
  {Pepper}, {De Lee}, {Collins}, {Latham}, {Muirhead}, {Chittidi},
  {Rojas-Ayala}, {Fleming}, {Rose}, {Tenenbaum}, {Ting}, {Kane}, {Barclay},
  {Bean}, {Brassuer}, {Charbonneau}, {Ge}, {Lissauer}, {Mann}, {McLean},
  {Mullally}, {Narita}, {Plavchan}, {Ricker}, {Sasselov}, {Seager}, {Sharma},
  {Shiao}, {Sozzetti}, {Stello}, {Vanderspek}, {Wallace}, \&
  {Winn}}]{Stassun19}
{Stassun}, K.~G., {et~al.} 2019, \aj, 158, 138

\bibitem[{{Su{\'a}rez Mascare{\~n}o} {et~al.}(2016){Su{\'a}rez Mascare{\~n}o},
  {Rebolo}, \& {Gonz{\'a}lez Hern{\'a}ndez}}]{Suarez16}
{Su{\'a}rez Mascare{\~n}o}, A., {Rebolo}, R., \& {Gonz{\'a}lez Hern{\'a}ndez},
  J.~I. 2016, \aap, 595, A12

\bibitem[{{Swift} {et~al.}(2015){Swift}, {Montet}, {Vanderburg}, {Morton},
  {Muirhead}, \& {Johnson}}]{Swift15}
{Swift}, J.~J., {Montet}, B.~T., {Vanderburg}, A., {Morton}, T., {Muirhead},
  P.~S., \& {Johnson}, J.~A. 2015, \apjs, 218, 26

\bibitem[{{Theano Development Team}(2016)}]{exoplanet:theano}
{Theano Development Team}. 2016, arXiv e-prints, abs/1605.02688

\bibitem[{{West} {et~al.}(2008){West}, {Hawley}, {Bochanski}, {Covey}, {Reid},
  {Dhital}, {Hilton}, \& {Masuda}}]{West08}
{West}, A.~A., {Hawley}, S.~L., {Bochanski}, J.~J., {Covey}, K.~R., {Reid},
  I.~N., {Dhital}, S., {Hilton}, E.~J., \& {Masuda}, M. 2008, \aj, 135, 785

\bibitem[{{West} {et~al.}(2015){West}, {Weisenburger}, {Irwin},
  {Berta-Thompson}, {Charbonneau}, {Dittmann}, \& {Pineda}}]{West15}
{West}, A.~A., {Weisenburger}, K.~L., {Irwin}, J., {Berta-Thompson}, Z.~K.,
  {Charbonneau}, D., {Dittmann}, J., \& {Pineda}, J.~S. 2015, \apj, 812, 3

\end{thebibliography}

\end{document}